\documentclass[aps,pra,preprint,groupedaddress,floatfix]{revtex4}

\usepackage{bm} 
\usepackage{graphicx}
\usepackage{epsfig}
\usepackage{subeqn}

\begin{document}

\title{Zero Sound in Dipolar Fermi Gases}

\author{Shai Ronen} 
\author{John L. Bohn}
\affiliation{JILA and Department of Physics, University of Colorado, Boulder, Colorado, USA}
\email{bohn@murphy.colorado.edu}
\date{\today}

\begin{abstract}

We study the propagation of sound in a homogeneous dipolar gas at zero
temperature, known as zero sound. We find that undamped sound
propagation is possible only in a range of solid angles around the
direction of polarization of the dipoles. Above a critical dipole
moment, we find an unstable mode, by which the gas collapses locally
perpendicular to the dipoles' direction.
\end{abstract}

\maketitle


\section{Introduction}

At ultracold temperatures, the usual propagation of thermodynamic
sound waves in a gas is suppressed due to the rarity of collisions.
Nevertheless, in a Fermi gas, collective excitations known as zero
sound can still propagate even at zero temperature
\cite{Landau57,LLSP2}.  These excitations involve a deformation of the
Fermi surface in a way distinguished from that of ordinary sound. The
theory of zero sound has been applied in the past to isotropic, two
component Fermi liquids. Depending on the properties and strength of
the interaction,one observes the possibility of different propagation
modes (longitudinal or transverse) as well as the possibility of no
zero sound propagation at all. Zero sound has been measured
experimentally in liquid He-3 \cite{Abel66,Roach76} in which both
longitudinal and transverse modes of zero sound co-exist. Another type
of collective excitations that exist for two component Fermi liquids
is spin-dependent vibrations known as spin waves.

Here we shall be interested in zero sound in a homogeneous,
\textit{single} component dipolar Fermi gas, where all the dipoles are
oriented in one direction by an external field. In particular we are
interested in the question of how zero sound propagation is affected
by the long range and anisotropic nature of the dipolar interaction.
In the case of Bose-Einstein condensates (BEC's), strong dipolar
effects have been observed in a gas of Chromium-52 atoms with
relatively big magnetic moments \cite{Lahaye07,Koch08,Lahaye08} - for
a recent review of dipolar BEC's see \cite{Lahaye09}. In the case of
Fermi gases, the atomic magnetic dipolar interaction is small compared
to the Fermi energy, so it may be hard to observe dipolar
effects. However much stronger effects are expected in an ultracold
gas of hetronuclear molecules with electric dipoles, where the dipolar
interaction is considerably larger (such an ultracold gas of fermionic
KRb molecules has been recently realized in experiment
\cite{Ospelkaus08a,Ni08}). Previous theoretical work on ultracold dipolar Fermi
gases studied their ground state properties and expansion dynamics in
the normal phase \cite{Goral01,Miyakawa08,Sogo08,Fregoso09}, as well
as BCS superfluidity \cite{Baranov02b,Baranov04}.  It has been shown
that both the critical temperature for the superfluid transition, and
the BCS order parameter, are sensitive to the aspect ratio of the
trap. In the case of a cigar trap, the order parameter becomes
non-monotonic as a function of distance from the center of the trap,
and even switches sign. However, zero sound refers to phenomena in the
normal phase, therefore we are interested in the regime of
temperatures which are well below the Fermi temperature $T_F$, but
still above the critical BCS temperature. Zero sound in Fermi dilute gases of alkali type
(with contact interactions), in the crossover from 3D to 2D, has been
recently studied in Ref. \cite{Mazzarella09}.

An interesting result due to Miyakawa eta al. (\cite{Miyakawa08},
\cite{Sogo08}) is that the Fermi surface of a dipolar gas is
deformed. They postulated an ellipsoidal variational \textit{ansatz}
for the Fermi surface, and by minimizing the total energy of the
system, found that that it is deformed into a prolate ellipsoid in the
direction of the polarization of the dipoles. As a preliminary to
studying zero sound, one of our aims in the present work is to compute
numerically the equilibrium Fermi surface, and thus, along the way,
also check the reliability of the variational method.

This paper is organized as follows. In section~(\ref{sec:shape}) we
find numerically the shape of the Fermi surface of a dipolar Fermi gas
at equilibrium, and compare it with the variational method. In section
~(\ref{sec:sound}) we study the zero sound in this gas, and how it
depends on the direction of propagation and the strength of the
interaction. We also find the instability of the gas above a critical
dipolar interaction strength, manifested in the appearance of a complex
eigenfrequency.  Finally we present our conclusions in
section~(\ref{sec:conc}).

\section{Shape of the Fermi surface \label{sec:shape}}

We consider a homogeneous, single component, dipolar gas of fermions
with mass $m$ and magnetic or electric dipole moment $d$. The dipoles
are assumed to be polarized along the $z$-axis. It is significant that
dipolar fermions continue to interact in the zero-temperature limit,
even if they are in identical internal states. This is a consequence
of the long-range dipolar interaction. The system is described by the
Hamiltonian:

\begin{eqnarray}
H=\sum_{i=1}^{N} -\frac{\hbar^2}{2 m}\nabla_i^2 +\sum_{i\neq j}V_{dd}(\bm{r}_i-\bm{r}_j),
\end{eqnarray}
where $V_{dd}(\bm{r})=(d^2/r^3)(1-3z^2/r^2)$ is the two-body dipolar interaction.
We use the  semi-classical approach in which the one-body density matrix is given by \cite{Miyakawa08}
\begin{eqnarray}
\rho(r,r')=\int \frac{d^3 k} {(2\pi)^3}
f\left(\frac{\bm{r}+\bm{r'}}{2},\bm{k}\right) e^ {i
  \bm{k}\cdot(\bm{r}-\bm{r'})},
\end{eqnarray}
where $f(\bm{r},\bm{k})$ is the Wigner distribution
function. The number density distributions in real and momentum space
are given by

\begin{eqnarray}
n(\bm{r})=\rho(\bm{r},\bm{r})=\int \frac{ d^3 k}{ (2 \pi)^3} f(\bm{r},\bm{k}) \\ \nonumber
\tilde{n}(\bm{k})=(2 \pi)^{-3} \int d^3 r f(\bm{r},\bm{k}).
\end{eqnarray}

Within the Thomas Fermi-Dirac approximation, the total energy of the system is given by
$E=E_{kin}+E_{d}+E_{ex}$, the sum of kinetic, dipolar and exchange energies, where:
\begin{eqnarray}
E_{kin}=\int \frac{d^3 k}{(2 \pi)^3}\frac{\hbar^2 k^2}{2 m} f(\bm{r},\bm{k}),
\end{eqnarray}

\begin{eqnarray}
E_{d}=\frac{1}{2}\int d^3 r \int d^3 r' V_{dd}(\bm{r}-\bm{r'})n(\bm{r}) n(\bm{r}'),
\end{eqnarray}

\begin{eqnarray}
\label{eq:Energy}
E_{ex}& =& -\frac{1}{2}\int d^3 r \int d^3 r' \int \frac{d^3 k}{(2 \pi)^3}\int \frac{d^3 k'}{(2 \pi )^3} 
V_{dd}(\bm{r-r'}) e^{i (\bm{k}-\bm{k'})\cdot (\bm{r}-\bm{r'})} \times \\ 
& & \nonumber f\left( \frac{\bm{r}+\bm{r'}}{2},\bm{k}\right) 
f \left( \frac{\bm{r+r'}}{2},\bm{k'} \right).
\end{eqnarray}

We shall consider a homogeneous system of volume $V$ with number
density $n_f$. Let $k_F$ be the Fermi wave number of an ideal Fermi
gas with that density. Then $n_f$ is related to $k_F$ by $n_f=k_F^3/(6
\pi^2)$.  For an homogeneous gas the exchange energy can be rewritten
as
\begin{eqnarray}
\label{eq:LDA}
E_{ex}= -V/2 \int \frac{d^3 k}{ (2 \pi)^3} \int \frac{d^3 k'}{(2 \pi)^3} f(\bm{k})f(\bm{k'})\tilde{V}_{dd}  (\bm{k}-\bm{k'}),
\end{eqnarray}
Here we have used the Fourier transform of the dipolar potential,
$\tilde{V}_{dd}(\bm{q})=(4 \pi/3)d^2(3 \cos^2 \theta_{\bm{q}}-1)$ where
$\theta_{\bm{q}}$ is the angle between the momentum $\bm{q}$ and the
direction of polarization, which is chosen to be along the $k_z$-axis.

For a homogeneous gas, the distribution function $f$ is a function of
$\bm{k}$ only. For an ideal gas or for a gas with isotropic
interactions at zero temperature, it is given by
$f(\bm{k})=\Theta(k_F-k)$, where $\Theta(k)$ is Heaviside's step
function. This describes a Fermi sphere with radius $k_F$.  In the
case of dipolar gas, Miyakawa et al.(\cite{Miyakawa08},
s\cite{Sogo08}) postulated the following variational
\textit{ansatz}:
\begin{eqnarray}
\label{var}
f(\bm{k})=\Theta\left( k_F^2-\frac{1}{\beta}(k_x^2+k_y^2)-\beta^2
k_z^2\right),
\end{eqnarray}
It is then possible to determine variationally the parameter $\beta$
that minimizes the total energy of the system. In general it is found
that $\beta<1$, namely the Fermi surface is deformed into a prolate
spheroid. This occurs due to the exchange interaction
(Eq.~(\ref{eq:LDA})) being negative along the direction of
polarization.

We shall first be interested here in finding the accuracy of this
variational method, and also  derive some analytical results for
the exchange energy, in a similar fashion to the well known exchange
energy of a homogeneous electron gas.

Instead of directly minimizing the total energy, our starting point is
that the quasi-particle energy $\epsilon(\bm{k})\equiv \frac{(2 \pi)^3}{V}
\frac{\delta E[f]}{\delta f} $ on the Fermi surface (i.e, the chemical
potential) must be constant in equilibrium. The quasi-particle energy is given by:
\begin{eqnarray}
\label{qp}
\epsilon(\bm{k})=\frac{\hbar^2 k^2}{2 m}-\int \frac{d^3 k'}{(2 \pi)^3}
f(\bm{k'}) \tilde{V}_{dd} (\bm{k}-\bm{k'}).
\end{eqnarray}
Further we shall assume that $f(\bm{k})$ is either 0 or 1, that is,
there is a well defined and sharp Fermi surface.  Our first
calculation is perturbative. For small $d$, the solution
of $\epsilon(\bm{k})=$constant, is found. Let $\bm{\hat{n}}$  be a unit direction vector in k-space.
Let $\bm{k}=k \bm{\hat{n}}$ lie on the (deformed) Fermi surface. Then we have:
\begin{eqnarray}
\bm{k}=\bm{\hat{n}} \left( k_F+\frac{m}{k_F \hbar^2} \int \frac{d^3 k'}{(2
\pi)^3} f(\bm{k'}) \tilde{V}_{dd} (k_F \bm{\hat{n}}-\bm{k'}) \right),
\end{eqnarray}
where, for weak interaction,  $f$ on the right side can be taken to be the distribution function of an ideal
gas. The integral can be evaluated analytically by using the convolution theorem, with the result:

\begin{eqnarray}
\label{pertshape}
k(\theta)=k_F+\frac{1}{9 \pi}\frac{m d^2 k_F^2}{\hbar^2} \left( 3 \cos^2(\theta)-1 \right),
\end{eqnarray}
where $\theta$ is the angle between $\bm{\hat{n}}$ and the $z$-axis.
For small $d$, Eq.~(\ref{pertshape}) is consistent with a Fermi
surface of slightly ellipsoidal shape, Eq.~(\ref{var}), with
$\beta=1-\frac{2}{9 \pi}\frac{m d^2 k_F}{\hbar^2}$. This confirms that
the ellipsoidal \textit{ansatz} is indeed correct for small $d$. The
corresponding energy is then 
\begin{eqnarray}
E/V=\frac{\hbar^2 k_F^5}{m} \left(
\frac{1}{20 \pi^2}-\frac{1}{405 \pi^4}(\frac{m d^2 k_F}{\hbar^2})^2
\right),
\end{eqnarray}
and the chemical potential is
\begin{eqnarray}
\mu=\frac{\hbar^2 k_F^2}{2 m}
\left( 1-\frac{28}{405 \pi^2}(\frac{m d^2 k_F}{\hbar^2})^2 \right).
\end{eqnarray}

Next, for arbitrary $d$, we solve numerically for
$\epsilon(\bm{k})=$constant. The calculation is a self consistent
iterative generalization of the perturbative calculation above, where
the constant is adjusted to obtain a fixed number of particles.

\begin{figure}

\resizebox{4in}{!}{\includegraphics{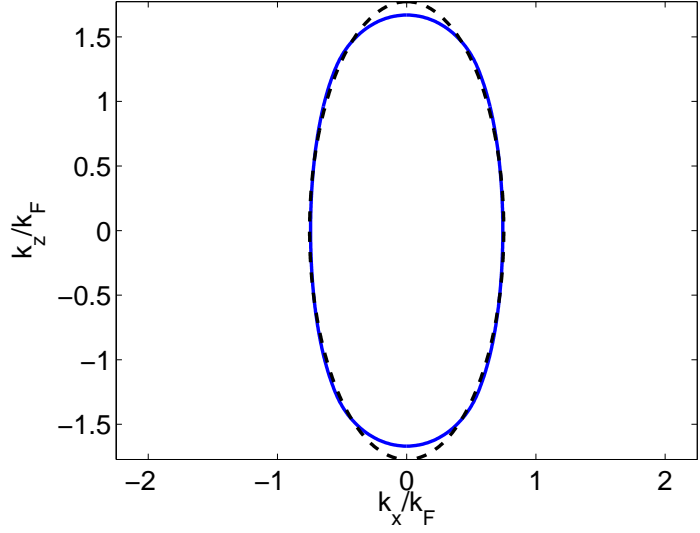}}
\caption{Shape of the Fermi surface of a homogeneous dipolar gas with
interaction strength $D=12.0$. Solid line: numerical
calculation. Dashed line: variational method \label{fig:shape}}
\end{figure}

It is convenient to present our results in terms of a dimensionless
parameter 
\begin{eqnarray}
D\equiv \frac{m d^2}{\hbar^2} k_F, 
\label{eq:dlength}
\end{eqnarray}
which signifies the strength of the dipolar interaction between two
particles separated by a distance $1/k_F$, in units of the Fermi
energy. This dimensionless expression appears in the equations for
$E/V$ and $\mu$ above. For reference, a gas density of $10^{12}$
cm$^{-3}$ with dipole moment of 1 Debye, and atomic mass of 100 amu,
has D=5.8. In Fig.~(\ref{fig:shape}) we compare the Fermi surface
found numerically with the variational Fermi surface for a strongly
interacting dipolar gas with $D=12.0$. We see that even for this
relatively strong interaction, the variational method works quite
well. The actual Fermi surface is not exactly an ellipsoid and is
slightly less prolate.

\begin{figure}
\resizebox{5in}{!}{\includegraphics{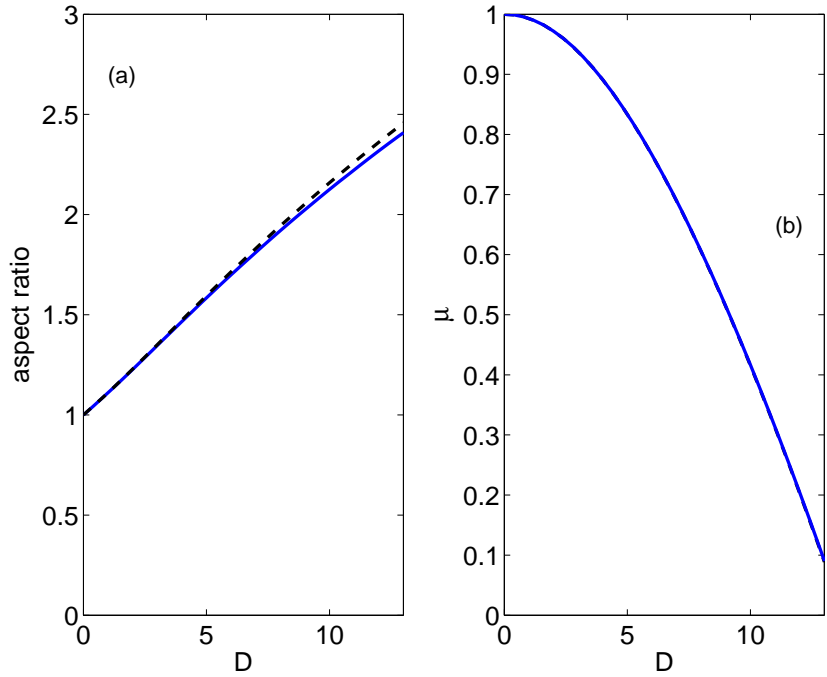}}
\caption{ a) Aspect ratio of the Fermi surface of dipolar gas as a function of the interaction strength $D$. Solid line: numerical
calculation. Dashed line: variational method.  
b) Chemical potential of dipolar gas as a function of the interaction strength $D$. $\mu$ is given in units
of $\hbar^2 k_F^2/m$. The numerical (solid) and variational (dashed) curves are virtually indistinguishable.
\label{fig:aspect}} 
\end{figure}

In Fig.~(\ref{fig:aspect}a) we compare the exact and variational 
results for the aspect ratio of the Fermi surface shape, and in
Fig.~(\ref{fig:aspect}b) we compare the chemical potentials. For both
quantities the variational method generally works very well.

Since our numerical results are close to those obtained by the
variational method, we might also expect an instability, due to the
inverse compressibility becoming negative, to occur around $D \approx
12.5$, as found in Ref.~\cite{Sogo08}. However, a cautionary note is in place, since for
such strong interactions one can expect deviations from the mean
field theory, which might stabilize the gas - as occurs in  the case of a
two-component Fermi gas with negative contact interactions \cite{Giorgini08}.

Recently, it has been suggested \cite{Fregoso09} that for strong
enough dipolar interactions, bi-axial nematic phase should
appear. Namely, above a critical interaction strength, the Fermi
surface will distort into an ellipsoid with three different semi-axis,
spontaneously breaking cylindrical symmetry. In our work, we assumed
from the start that cylindrical symmetry of the equilibrium state is not broken, However, the bi-axial phase was predicted to occur around $D
\approx 30$, which is  above the compressional instability limit of $D=9.5$. In what follows, we shall restrict ourselves
to discuss zero sound for $D<9.5$. In fact, we find below that the gas becomes unstable and collapses even below this limit.

\section{Zero Sound \label{sec:sound}}

We now consider collective excitations of a homogeneous dipolar Fermi
gas at zero temperature, i.e zero sound.  The normal speed of
thermodynamic sound (first sound) is $c^2=\frac{1}{m}\frac{\partial
  P}{\partial n}$ where $P=-\frac{\partial E}{\partial V}$ is the
pressure. However, thermodynamic sound is highly attenuated for a
rarefied gas at low temperatures (in the collisionless regime). The
attenuation of sound is due to transfer of energy from the sound wave
to random molecular motion (heat), and depends on the ratio of the
collision rate to the sound frequency. Specifically, the collisionless
regime is obtained for $\omega \tau>>1$, where $\tau$ is the mean time
between collisions, and $\omega$ the frequency of sound.
In particular, for a Fermi gas, the collisionless regime is rapidly obtained  when the
temperature drops below the Fermi temperature, and collisions are quenched by Fermi statistics.  
However Landau \cite{Landau57}
discovered that a collective excitation of a Fermi gas at zero
temperature is still possible in this case. This zero sound is
essentially a propagation of a deformation of the Fermi surface.  The
theory of zero sound as applied to a two component Fermi gas with
short range interactions is described in standard textbooks, e.g
\cite{LLSP2,QMPS}.

Our goal is to apply this theory to the case of the one component
dipolar gas. The local quasi-particle energy at position $\bm{r}$
is given by
\begin{eqnarray}
\label{qp2}
\epsilon( \bm{r}, \bm{k}) &\equiv& \frac{(2 \pi)^3} {V} \frac {\delta E} {\delta f(\bm{r},\bm{k})}= \\  \nonumber 
& & =\frac {\hbar^2 k^2} {2 m} +\int d^3 r' \frac {d^3 k'} {(2 \pi)^3} V_{dd}(\bm{r}-\bm{r'}) f(\bm{r'},\bm{k'})-\int 
\frac{d^3 k'} {(2 \pi)^3} f(\bm{r},\bm{k'}) \tilde{V}_{dd}(\bm{k'}-\bm{k}),
\end{eqnarray}
where the last term, the contribution of exchange interaction, has
been calculated within local density approximation (as in
Eq.~(\ref{qp})), suitable for collective excitations with long
wavelength.  Note that, compared to Eq.~(\ref{qp}), Eq.~(\ref{qp2}) shows
an extra direct-interaction term involving $V_{dd}(\bm{r}-\bm{r'})$.
For a spatially homogeneous distribution function this term vanishes,
because the angular average of the dipolar interaction is zero.

We apply a Boltzmann transport equation which reads:

\begin{eqnarray}
\label{eqmotion}
\frac{d f}{ d t}\equiv \frac{\partial f}{ \partial t}+\frac{1}{\hbar}\nabla_{\bm{r}} f  \cdot \nabla_{\bm{k}} \epsilon - 
\frac{1}{\hbar}\nabla_{\bm{k}} f  \cdot \nabla_{\bm{r}} \epsilon = I(f),
\end{eqnarray}
where $I(f)$ is the collision integral, and $\epsilon$ can be taken,
to first order, to be the equilibrium quasi-particle energy.  The
collisional integral can be neglected in the collisionless regime we
are interested in here.  However, more generally, it is interesting to
note that the dipolar interaction is long range, giving rise to both a
mean field interaction ($V_{dd}(\bm{r}-\bm{r'})$ term in the right side of
Eq.~(\ref{qp2})) and a collisional cross section that enters
$I(f)$. In this respect the dynamics of dipolar Fermi gas are an
intermediate between, on the one hand, plasma dynamics which to a first approximation
are controlled by the mean field potential (as in Vlasov equation), and
on the other hand, the dynamics of Fermi liquids with short range interactions, which are
controlled by the local exchange interaction and the collisional
integral.

For small deviations from  equilibrium, the distribution function
changes  in the near vicinity of the equilibrium Fermi Surface,
and can be written in the form

\begin{eqnarray}
f=f_{e}+e^{i (\bm{q} \cdot \bm{r}- \omega t)} \nu(\bm{k})
\delta(\epsilon_{e}(\bm{k})-\mu),
\label{eq:ansatz}
\end{eqnarray}
where $\mu$ is the equilibrium chemical potential, and $\epsilon_{e}$
is as in Eq.~(\ref{qp}). $\nu(\bm{k})$ is a function defined on the
equilibrium Fermi Surface, and will signify the eigenmode of the zero
sound wave. This perturbation describes a mode with spatial
wavenumber $\bm{q}$ and frequency $\omega$.

Eqs.~(\ref{eqmotion}) and (\ref{qp2}) then give:
\begin{eqnarray}
\label{eq:zs1}
\left(\omega -\bm{v}(\bm{k}) \cdot \bm{q} \right) \nu(\bm{k})= -\bm{v}(\bm{k})
\cdot \bm{q} \oint dS' \frac {\tilde{V}_{dd}(\bm{k}-\bm{k'})
\nu(\bm{k'})}{(2\pi)^3 \hbar |\bm{v}(\bm{k'})|} + \\ \nonumber + \bm{v}(\bm{k}) \cdot
\bm{q} \tilde{V}_{dd}(\bm{q}) \oint dS'
\frac{\nu(\bm{k'})}{(2 \pi)^3 \hbar |\bm{v}(\bm{k'})|}.
\end{eqnarray}
where $\bm{v}(\bm{k}) \equiv \frac{\nabla_{\bm{k}}
\epsilon_{e}}{\hbar}$. In the above equation, 
$\bm{k}$ and $\bm{k'}$ lie on the equilibrium Fermi surface. Also the
integrals are taken on this surface. The first integral is an exchange
interaction term, and the second a direct term. They can be combined to
obtain:
\begin{eqnarray}
\label{eq:zs}
\left(\omega -\bm{v}(\bm{k}) \cdot \bm{q} \right) \nu(\bm{k}) = \bm{v}(\bm{k})
\cdot \bm{q} \oint dS' \frac { \left( \tilde{V}_{dd}(\bm{q})-\tilde{V}_{dd}(\bm{k}-\bm{k'}) \right)
\nu(\bm{k'})}{(2\pi)^3 \hbar |\bm{v}(\bm{k'})|}.
\end{eqnarray}

The problem of zero sound of a dipolar gas is thus reduced to finding,
for a given wavenumber $\bm{q}$, the eigenvalues $\omega$ and
eigenmodes $\nu(\bm{k})$. In fact, it can be seen from
Eq.~(\ref{eq:zs}) that $\omega$ is linear in $|\bm{q}|$, and thus we
expect a linear phonon spectrum with a speed of sound which in general
depends on the relative direction of the propagation direction
$\bm{q}$ with respect to the direction of polarization. It can be seen
that for undamped vibrations, the speed of sound
$s(\hat{\bm{q}})=\frac{\omega}{|\bm{q}|}$ must exceed a critical
velocity $v_{\mathrm{damp}}(\hat{\bm{q}}) \equiv \max_{\bm{k}} \{ \bm{v}(\bm{k}) \cdot \hat{\bm{q}} \}$, 
where $\hat{\bm{q}}=\bm{q}/|\bm{q}|$. To see this,
replace $\nu$ by another unknown function
$\tilde{\nu}(\bm{k})=(\omega-\bm{v}({\bm{k}})\cdot\bm{q})\nu(\bm{k})$. Then
Eq.~(\ref{eq:zs}) can be rewritten as:
\begin{eqnarray}
\tilde{\nu}(\bm{k})= 
\bm{v}(\bm{k}) \cdot \bm{q} \oint dS' \frac{
\left( \tilde{V}_{dd}(\bm{q})-\tilde{V}_{dd}(\bm{k}-\bm{k'}) \right) \tilde{\nu}(\bm{k'})} {(\omega
-\bm{v}(\bm{k'}) \cdot \bm{q})(2\pi)^3 \hbar |\bm{v}(\bm{k'})|}
\end{eqnarray}
When $\omega< \max_{\bm{k}} \{ \bm{v}(\bm{k}) \cdot \bm{q} \}$, the
integrand has a pole, which must be avoided by going around it in the
complex plane, giving an imaginary part which signifies the decay of
such a vibrational mode.

Eq.~(\ref{eq:zs}) can be solved numerically by discretizing it on the
Fermi surface. We use an evenly spaced, 64x64 grid, in spherical
coordinates $\theta,\phi$, where the 'north pole' is in the direction
of polarization of the dipoles. In accordance with the discussion
above, we select the undamped modes which satisfy
$s(\hat{\bm{q})}>v_{\mathrm{damp}}(\hat{\bm{q}})$. For that purpose,
it is sufficient to look for a few eigenmodes with the largest
eigenvalues, using an Arnoldi method. We checked our numerical code by
calculating the modes for the case of an isotropic Fermi Fluid
\cite{LLSP2}, and comparing to the known analytic solutions. 

Before discussing our results for the dipolar gas, it will be helpful
to briefly review the known results for two component isotropic Fermi
liquids.  The case relevant to us is that of same spin vibrations, in
which the two components move in unison. For an isotropic fluid the
Fermi surface is a sphere. The interaction term
\begin{eqnarray}
F(\bm{q},\bm{k},\bm{k'})\equiv
\tilde{V}_{dd}(\bm{q})-\tilde{V}_{dd}(\bm{k}-\bm{k'}),
\label{eq:F}
\end{eqnarray}
which appears in Eq.(\ref{eq:zs}), is replaced, in the case of
isotropic liquids, by $F(\bm{k},\bm{k'})= \frac{2 \pi^2 \hbar}{m k_F}
\left( F_0+F_1 \bm{\hat{k}} \cdot \bm{\hat{k}'}\right)$, where $F_0$
and $F_1$ are the Landau parameters.  One finds the
following. When $F_1=0$ and $F_0>0$ there is a longitudinal mode which
is concentrated on the Fermi sphere around the forward direction of
propagation, which becomes more and more concentrated around that
point as $F_0 \rightarrow 0$. For $F_1=0$ and $F_0<0$, no zero sound
propagation exists.  For $F_1>6$, there also exists a transverse mode
which has a vortex like structure on the Fermi surface around the
direction of propagation.

\begin{figure}
\resizebox{4in}{!}{\includegraphics{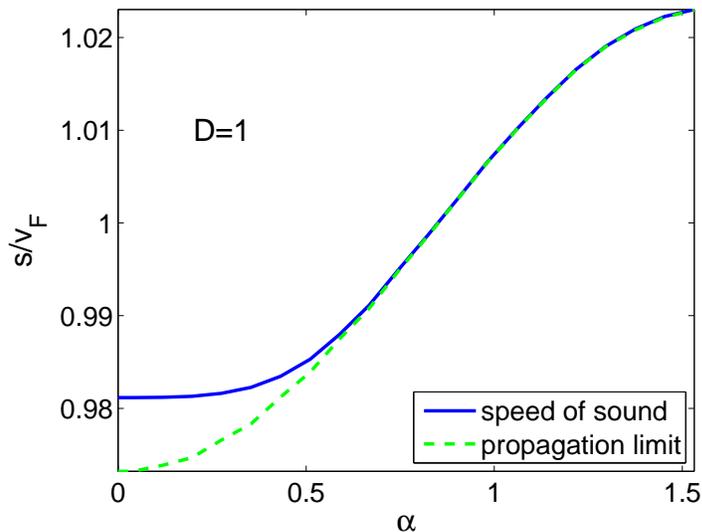}}
\caption{Speed of sound (solid curve) in a dipolar Fermi gas as a
function of the angle of propagation $\alpha$ relative to the
direction of polarization, for dipolar interaction strength $D=1$. The
speed is measured in units of $v_F \equiv \hbar k_F/m$. The dashed
curve represents a lower bound on the speed of any undamped mode. When
the two curves merge, zero sound modes are damped and fail to
propagate.
\label{fig:speed1}}
\end{figure}

We now examine the results for the dipolar
gas. Fig.~(\ref{fig:speed1}) shows the speed of sound $s=\omega/q$ in
the dipolar Fermi gas for interaction strength parameter $D=1$, as a
function of the angle $\alpha$ between the direction of sound
propagation and the direction of polarization. We find one zero sound
mode which, for small $\alpha$, satisfies the criterion for
propagation However sound does not propagate in all directions: for
$\alpha> \alpha_c \approx 0.6$ the gap between the calculated speed of
sound and the propagation limit $v_{\mathrm{damp}}$ is essentially zero. This
means that there is no undamped propagating zero sound for a wide
range of angles around the direction perpendicular to the direction of
polarization.

To understand this result, we borrow lessons learned from the case of
isotropic Fermi fluids. There, we know that spin-independent, longitudinal zero sound
mode, only propagates for repulsive interaction (positive landau
parameter $F_0$), and is damped for attractive interaction (negative
landau parameter $F_0$).  Thus, it makes sense that the anisotropy of
the dipolar interaction gives rise to sound propagation only in a
certain range of directions around the direction of polarization,
where the effective interaction in momentum space is repulsive. 

For weak interactions, we can gain some insight into the observed
behavior, as follows. In this case, the Fermi surface is nearly a
sphere. Moreover, the eigenmode $\nu(\bm{k})$ is appreciably different
from zero only in a small region on the Fermi surface around the
direction of $\bm{q}$.  This behavior is expected from the general
Landau theory \cite{LLSP2} and is confirmed by examination of our
numerical solutions.  Therefore, we may replace, in Eq.~(\ref{eq:zs}),
the interaction term $F(\bm{q},\bm{k},\bm{k'})$
with its value for $\bm{k'},\bm{k}$ close to $\hat{\bm{q}}$. Since
$\tilde{V}_{dd}$ is anisotropic, $F$ does not have in general a
single-valued limit as $\bm{k},\bm{k'} \rightarrow \bm{q}$, because
the limit depends on the direction of $\bm{k'}-\bm{k}$. Nevertheless
we can observe the following. For the case $\alpha=0$, i.e, $\bm{q}$
along the direction of polarization, we find $F$ does have the single
valued limit of $F=4\pi d^2$. This is completely analogous to a
positive Landau parameter $F_0$ in the isotropic Fermi liquid 
\cite{LLSP2}, and thus we obtain a propagating zero sound mode in this
direction. On the other hand, for $\alpha=\pi/2$ it is easy to
establish an upper bound $F(\bm{\hat{q}},\bm{k},\bm{k'}) \leq
0$. Although here $F$ does not have a single-valued limit, we can
still expect, from analogy to the $F_0<0$ case in the isotropic Fermi liquid, that longitudinal zero mode does not propagate
for this case.

As observed above, numerically we determine the critical angle for
propagation to be around $\alpha=0.6$ for $D=1$. However we caution
that we could not get to limit of very weak interactions $D\rightarrow
0$ by our numerical method, since in this case the eigenmodes tend to
be highly concentrated around the forward direction of propagation,
and cannot be resolved with our grid. 

\begin{figure}
\resizebox{4in}{!}{\includegraphics{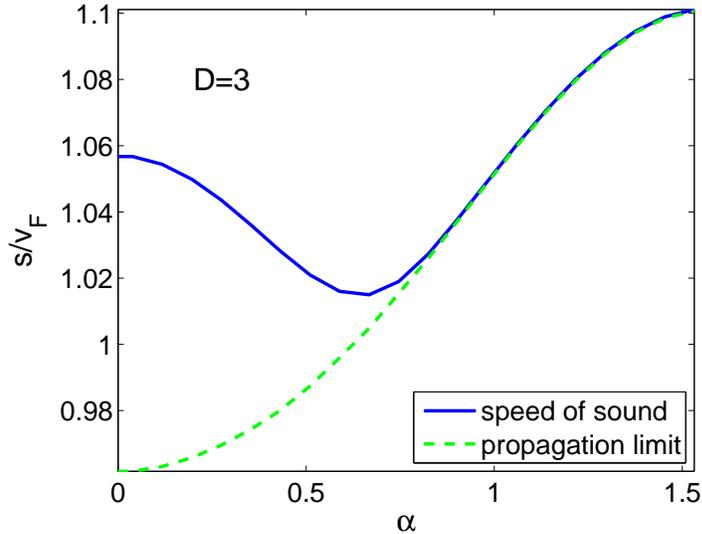}}
\caption{Speed of sound in a dipolar Fermi gas as a function of the
angle of propagation $\alpha$, for $D=3$. \label{fig:speed3}}
\end{figure}

As we increase the interaction strength from $D=1$ to $D=3$
(Fig.~(\ref{fig:speed3})), there is a noticeable change in the shape
of the speed of sound curve. Yet, we still have one propagating mode
with some critical angle beyond which zero sound does not propagate.

The deformation of the Fermi surface studied in the previous section
contributes somewhat  to the change in the speed of sound. It
somewhat changes the Fermi surface element and the Fermi velocity
$\nu({\bm{k}})$ in Eq~(\ref{eq:zs}).  However, the main factor giving
rise to a change from propagation to damping of zero sound as a
function of direction is the anisotropy of the dipolar interaction
which  directly enters that equation.

For $D>4$ we find that the solution of Eq.(~\ref{eq:zs}) admits
complex eigenvalues, in particular an eigenvalue which is purely
imaginary and lies on the upper complex plane.  This signifies an
instability, since according to Eq.~(\ref{eq:ansatz}) such a mode
grows exponentially. The imaginary eigenvalue appears first about
$D=4.0$ and $\alpha=\pi /2 $. Thus, the gas is unstable to growth of
density waves in the plan perpendicular to the direction of the
dipoles. For $D>4$ we find also additional propagating modes (with
real valued frequncies), in particular a transverse mode analog to the
transverse mode in superfluid He-3. However, because of the
instability, we conclude that these modes could not be experimentally
observed, at least unless some additional stabilizing mechanism is introduced.

As noted in the previous section, the compressibility of the gas only
becomes infinite at $D=9.5$.  However it is physically reasonable that
a dipolar gas becomes unstable prior to that. The reason is that the
compressibility is a derivative of the pressure, which is an isotropic
quantity. But due to the anisotropic interaction, the gas is actually
more sensitive.to collapse in a specific direction, i.e, perpendicular
to the dipoles. This tends toward the growth of over-density regions
where the dipoles are oriented head to tail, lowering the potential
energy.

This mechanism is similar to that already long known for homogeneous
dipolar Bose gases.  A homogeneous Bose gas with a purely dipolar
interaction is unstable, but it can be stabilized by the addition of
repulsive short range interaction with scattering length $a>0$
\cite{ODell04}. In terms of a dimensionless dipolar interaction
strength $D_B=m d^2/\hbar^2 a$, the stability criterion of the
homogeneous Bose gas is $D_B<3$. Also, from examining the dispersion
relation for the Bose gas it can be shown that it becomes first
unstable at $D_B=3$ due to density waves perpendicular to the
direction of polarization. Finally, we note that $D_B$ and the dipolar
Fermi parameter $D$ (Eq.~(\ref{eq:dlength}) are related by the simple
exchange of the Fermi length scale $1/k_F$ with the scattering length
$a$. It is interesting to observe that in these respective units, the
criteria for the onset of instability in Bose and Fermi gases are very
similar in magnitude.

{\it Experimental prospects:} We outline some additional
considerations regarding the experimental prospects of observing zero
sound in dipolar Fermi gas. The condition to have a normal phase is
that $T>T_c$, where $T_c$, the critical temperature for the superfluid
transition, is given by $T_c \approx 1.44 T_f \exp \left( -\pi^3/(4 D)
\right)$, with $T_f$ the Fermi temperature \cite{Baranov02b}. At the
same time, zero sound attenuation is proportional to $T^2/\tau$, where
$\tau$ is the relaxation time due to collisions
\cite{Abrikosov59}. This will set some maximal temperature $T_m$ for
realistic experimental detection of zero sound. Determining the zero
sound attenuation at finite temperature theoretically is a task beyond
the scope of this paper , but we note the following: observing zero
sound in the normal phase requires $T_m>T>T_c$, and we thus obtain the
condition $D<\pi^3/ \left(4 \log (1.44 T_f/T_m) \right)$.  As an
illustration, if $T_m=0.1 T_f$, we obtain $D<2.9$. Still, even for
smaller $T_m$, restricting the observation of the zero sound to
smaller $D$, it should be possible to see the propagation limited to
certain directions only (as in Fig.~(\ref{fig:speed1}).  Finally, the
above discussion is based on the requirement that
$T>T_c$. Nevertheless, we note that, under certain conditions, zero
sound of the normal phase can still persist below $T_c$, into the
superfluid phase \cite{Legget66,Davis08}.

\section{Conclusions \label{sec:conc}}
In conclusion, we have studied numerically the deformation of the
Fermi surface and zero sound excitations in a homogeneous, single
component, degenerate Fermi gas of polar atoms or molecules polarized
by an external field. We find that the Fermi surface is described very
well by the variational \textit{ansatz} proposed in \cite{Miyakawa08}
up to about dipolar interactions strength $D \approx 12.5$, where a
compressional instability was predicted.  We then study zero sound in
the range $D=0-10$. For $D<4$ we find that zero sound can propagate
with a longitudinal mode in directions between $\alpha=0$ and some
critical $\alpha$ which depends weakly on $D$ but is generally close
to $0.6-0.7$ radians ($\alpha$ being the angle between the direction
of propagation and the direction of polarization). Beyond this angle,
undamped propagation of zero sound is not possible. For $D>4$, we find
a complex eigenfrequency indicating that the dipolar gas collapses due
to an unstable mode in the direction perpendicular to the dipoles.

{\it Note added}: after completion of this work, a work by Chan et
al. appeared \cite{Chan09}, who also study properties of dipolar Fermi gas,
including zero sound.




\begin{acknowledgments}
SR is grateful for helpful and motivating discussions with Jami
Kinnunen and Daw-Wei Wang. We acknowledge financial support from the NSF. 
\end{acknowledgments}

\section*{References}
\bibliographystyle{unsrt}
\bibliography{biblo}
\end{document}